\documentclass[aps,pre,reprint,superscriptaddress]{revtex4-2}

\usepackage{graphicx}
\graphicspath{ {./figures/} }
\usepackage{subfigure}

\begin{document}


\title{Density estimation for ordinal biological sequences and its applications}


\author{Wei-Chia Chen}
\email[Correspondence: ]{wcchen@ccu.edu.tw}
\affiliation{Department of Physics, National Chung Cheng University, Chiayi 62102, Taiwan, R.O.C.}

\author{Juannan Zhou}
\affiliation{Department of Biology, University of Florida, Gainesville, Florida 32611, U.S.A.}

\author{David M. McCandlish}
\affiliation{Simons Center of Quantitative Biology, Cold Spring Harbor Laboratory, Cold Spring Harbor, New York 11724, U.S.A.}


\date{\today}

\begin{abstract}
Biological sequences do not come at random. Instead, they appear with particular frequencies that reflect properties of the associated system or phenomenon. Knowing how biological sequences are distributed in sequence space is thus a natural first step toward understanding the underlying mechanisms. Here we propose a new method for inferring the probability distribution from which a sample of biological sequences were drawn for the case where the sequences are composed of elements that admit a natural ordering. Our method is based on Bayesian field theory, a physics-based machine learning approach, and can be regarded as a nonparametric extension of the traditional maximum entropy estimate. As an example, we use it to analyze the aneuploidy data pertaining to gliomas from The Cancer Genome Atlas project. In addition, we demonstrate two follow-up analyses that can be performed with the resulting probability distribution. One of them is to investigate the associations among the sequence sites. This provides us a way to infer the governing biological grammar. The other is to study the global geometry of the probability landscape, which allows us to look at the problem from an evolutionary point of view. It can be seen that this methodology enables us to learn from a sample of sequences about how a biological system or phenomenon in the real world works.
\end{abstract}


\maketitle


%
%
\section{Introduction \label{Introduction}}
The technology of DNA sequencing has evolved immensely since its invention by Sanger in 1977 \cite{Sanger1977}. Particularly, in the past twenty years we have seen the technology transforming from the first-generation electrophoretic sequencing into the second-generation massively parallel sequencing and the third-generation real-time, single-molecule sequencing. For a review, see reference \cite{Shendure2017}. Accompanying those advances are applications of DNA sequencing in new ways or to new areas, such as ChIP-seq \cite{Johnson2007} and RNA-seq \cite{Cloonan2008, Lister2008, Mortazavi2008, Nagalakshmi2008, Wilhelm2008}. Needless to say, a natural product of these sequencing experiments is a wealth of biological sequences.

It is now a well known fact that biological sequences do not come at random. For instance, DNAs in a segment of genome are usually arranged in some order, encoding a piece of genetic information. Thus, given a sample of biological sequences, a fundamental question is {\it what the underlying probability distribution from which the sequences were drawn looks like} \cite{Durbin1998, Schneidman2006, Humplik2017, Cocco2018, Riesselman2018}. In a recent work \cite{Chen2021} we developed a novel density estimation method for biological sequences based on Bayesian field theory \cite{Bialek1996, Nemenman2002, Enblin2009, Enblin2013, Kinney2014, Kinney2015, Chen2018}. This method, called SeqDEFT (standing for Sequence Density Estimation using Field Theory), was aimed for {\it categorical} sequences, that is, sequences made up of elements that are distinct from each other, but do not have a natural ordering. Typical examples of categorical sequence data include DNA sequences (consisting of the four nucleotides) and protein sequences (consisting of the twenty amino acids). The SeqDEFT method can be regarded as a nonparametric extension of the position weight matrix model \cite{Staden1984, Stormo2013} or the Potts model \cite{Lapedes1999, Yeo2004, Bialek2007, Weigt2009, Mora2010, Balakrishnan2011, Morcos2011, Ekeberg2013, vanNimwegen2016, Levy2017}, depending on the chosen value of a hyperparameter. We showed that SeqDEFT was able to provide a more detailed view of the probability landscape than these commonly used models in two different biological contexts. 

In this work we propose a new method, again within the framework of Bayesian field theory, for the density estimation on a different type of sequence data, whose elements do have a natural ordering. Such {\it ordinal} sequence data are not uncommon in biology. An example is the karyotype of a cell or an organism, which contains the copy number of each chromosome thereof. (In biology, a karyotype includes the sizes, numbers, and shapes of the chromosomes. Here we are adopting a simplified version of karyotype and only look at the copy numbers.) Thanks to the advancement of DNA sequencing technology, today a karyotype can even record the copy number of each gene in the whole genome \cite{Shendure2017}. The ability to obtain karyotypes either in the chromosome-level or in the gene-level has a huge impact on medicine, as it has been shown that patients having certain diseases tend to possess some particular patterns of chromosome or gene copy number variations (e.g., \cite{Taylor2018}). Such molecular information is playing a more and more important role in clinical studies. For an overview, see reference \cite{Nogrady2020}.

For instance, in the most recent WHO (World Health Organization) classification diffuse gliomas, which is the most common malignant tumor of the central nerve system in adults, have been classified into three subtypes, oligodendroglioma, astrocytoma, and glioblastoma, according to their molecular features as well as traditional histology \cite{Whitfield2022}. Both oligodendroglioma (WHO Grade 2, 3) and astrocytoma (WHO Grade 2, 3, 4) have the IDH gene mutant, but oligodendroglioma has the additional codeletion of chromosome arms 1p and 19q, whereas astrocytoma does not. On the other hand, glioblastoma (WHO Grade 4) does not have the IDH gene mutant, but many patients have gain in chromosome 7 and loss in chromosome 10, as well as other gene mutations. The 5-year survival rate of low-grade glioma is about 80$\%$, and for high-grade glioma it is below 5$\%$ \cite{Whitfield2022}. This refined classification, based on both molecular features and traditional histology, not only can better predict disease progression but may also suggest novel treatments for the disease \cite{Whitfield2022}.  

The density estimation method proposed here shares many properties with its categorical analogue (i.e., SeqDEFT). In particular, it includes the position weight matrix model and the Potts model as limiting cases. Also, being a nonparametric method, it can be expected to provide a better estimate of how ordinal sequences are distributed in sequence space than those traditional methods. Moreover, we propose two follow-up analyses that can be conducted with a probability distribution to extract more information from data. First, we show that by examining the associations among the sequence sites, we can find out which patterns dominate in a certain background. These ``rules'' are analogous to the language grammar that one has to obey in order to make valid words, and thus can be regarded as the biological grammar governing the system or phenomenon. Second, we can visualize a probability landscape using a dimensionality reduction technique based on an evolutionary model. Doing so allows us to imagine what a population is likely going to experience as it wanders around the sequence space. 

This article is organized as follows. In Section \ref{Formalism} we develop the density estimation method and point out the difference between the method and its categorical analogue. Then we perform density estimation, along with the two follow-up analyses, on the aneuploidy data pertaining to gliomas collected by The Cancer Genome Atlas (TCGA) project \cite{Hutter2018} in Section \ref{Result}. Finally, we conclude with a summary and discussion in Section \ref{Summary and Discussion}.

%
%
\section{Formalism \label{Formalism}}
In this section we develop our new method for density estimation on ordinal sequence data within the framework of Bayesian field theory. We call this method ordinal SeqDEFT. Although some of the results have been presented in the literature, especially \cite{Chen2021} and \cite{Kinney2015}, we include all the essential stuff to make the content of this section self-contained. 

Our goal here is to estimate a probability distribution with a sample of ordinal biological sequences. Assume that the sequences have length $\ell$ and each site of the sequences has $\alpha$ possible elements. The number of all possible combinations is equal to $\alpha^\ell$. Each combination is a possible sequence.

\begin{figure}[t]
\includegraphics[width=0.9\linewidth]{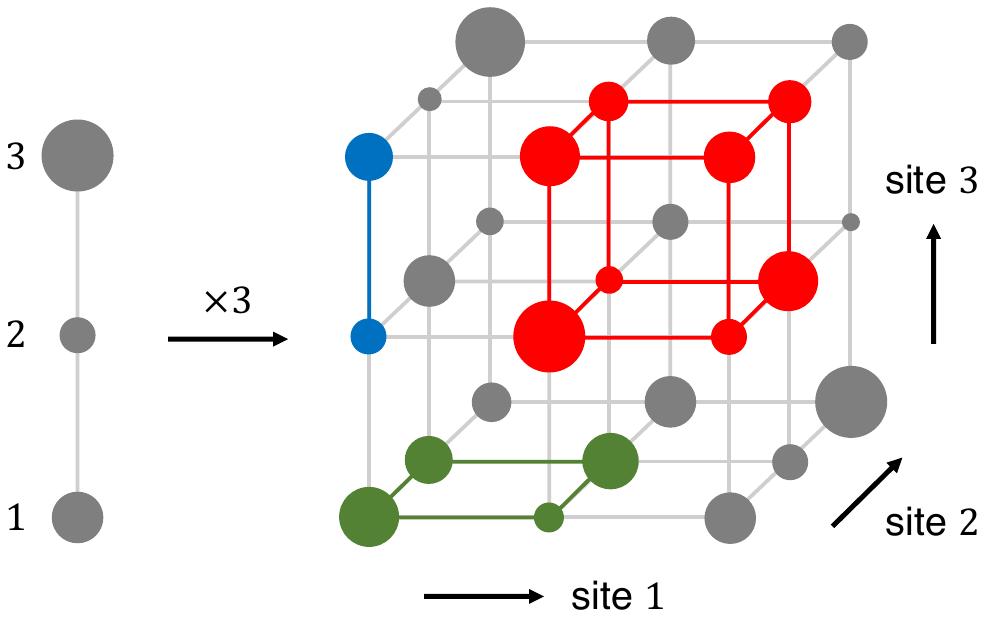}
\caption{Representation of an $\alpha^\ell = 3^3$ ordinal sequence space. In this example, each sequence consists of three sites, and each site can be in one of three states, say, 1, 2, 3. The order of the states is represented by a path graph of length 3 (left). The sequence space, in turn, is represented as a grid graph obtained by a 3-fold Cartesian product of the path graph (right). Each direction of the grid graph corresponds to a particular site. This graph has 27 nodes in total, each node represents a sequence and the size of the node is proportional to the probability of the sequence. The computation of $P$th-order associations involves probabilities of the sequences that form a $P$-dimensional hypercube. For $P = 1, 2, 3$, the hypercube is edge (blue), face (green), and cube (red), respectively. }
\label{fig:1}
\end{figure}

The major difficulty in analyzing sequence data arises from the lack of relationships among the sequences. Without knowing the sequences' relationships, it is not clear how to define meaningful quantities for the problem. Thus, in order to proceed, we need to introduce some kind of relationships among the sequences, and such relationships should be chosen according to the nature of the data in question. For instance, for categorical sequence data, each site is allowed to mutate from one element to all the other elements. Such a relationship can be represented by a {\it complete graph}, and the whole sequence space can then be represented by a Cartesian product of $\ell$ complete graphs, namely, a {\it Hamming graph} \cite{Chen2021}. For ordinal sequence data considered here, each site can only jump from one element to the element(s) next to it in order. This relationship can be represented by a {\it path graph}, and thus we can use an $\ell$-fold Cartesian product of path graphs, which results in a {\it grid graph}, to represent the sequence space. Fig.~\ref{fig:1} shows an example of this procedure with $\alpha^\ell = 3^3$. 

\begin{figure*}[t]
\includegraphics[width=0.98\linewidth]{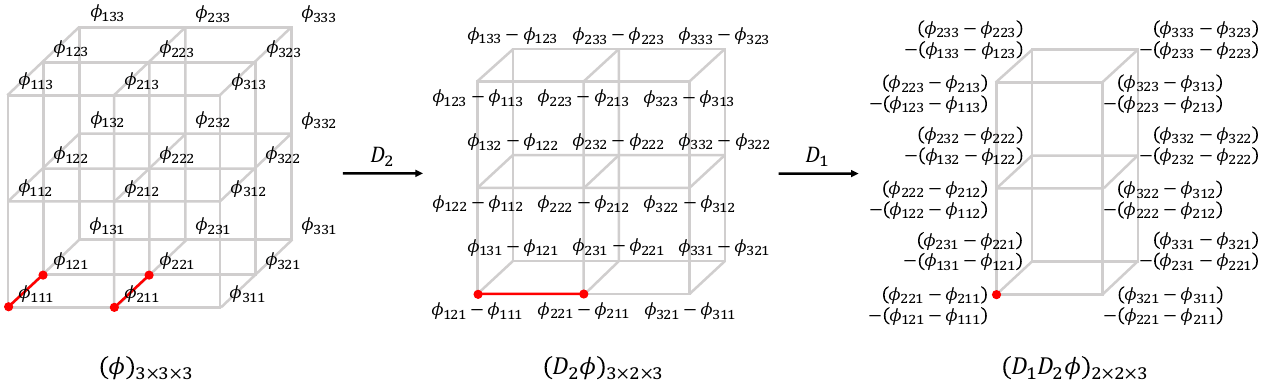}
\caption{Illustration of taking discrete differences of a field $\phi$ defined on an $\alpha^\ell = 3^3$ grid graph, as the one shown in Fig.~\ref{fig:1}. By construction, $\phi$ is a 3-dimensional array with 3 elements in each direction. Upon taking two consecutive discrete differences, first along the second direction ($D_2$) and then along the first direction ($D_1$), we obtain an array $D_1 D_2 \phi$ with dimension $2 \times 2 \times 3$. In each step, the components of the resulting array are the differences of the adjacent values of the previous array in that direction. Each component of the array $D_1 D_2 \phi$ involves four sequences that form a face in the sequence space.}
\label{fig:2}
\end{figure*}

Having represented the sequence space by a grid graph, we denote the probability for a sequence by $Q_{i_1 \cdots i_\ell}$ where the indices $i_1, \dots, i_\ell$, each corresponding to a site, are from $1$ to $\alpha$. We use this notation to emphasize that we are treating the probability distribution $Q$ as an $\ell$-dimensional array with $\alpha$ elements in each direction. Then we parametrize the distribution $Q$ by a field $\phi$ in the following manner:
\begin{equation} \label{eq:Q_phi}
Q_{i_1 \cdots i_\ell} = \frac{e^{-\phi_{i_1 \cdots i_\ell}}}{\sum_{j_1 \cdots j_\ell} e^{-\phi_{j_1 \cdots j_\ell}}}.
\end{equation}
Doing so, $Q$ automatically satisfies the conditions, $Q_{i_1 \cdots i_\ell} \ge 0$ and $\sum_{i_1 \cdots i_\ell} Q_{i_1 \cdots i_\ell} = 1$, required in order for it to be a valid probability distribution, whereas the field $\phi$ can take on any form or any value. 

To control the ``smoothness'' of $\phi$, that is, how $\phi$ changes from one sequence to another, we impose the following prior distribution on it:
\begin{equation}
p(\phi | a) \propto e^{-S_a^0[\phi]},
\end{equation}
where the functional $S_a^0[\phi]$, called the prior action, is given by
\begin{equation} \label{eq:prior_S}
S_a^0[\phi] = \frac{a}{2s} \ \! \phi^T \Delta^{(P)} \phi.
\end{equation}
Note that we have used the vector notation $\phi^T \Delta^{(P)} \phi$ to represent the operation acting on $\phi$ in order to make the formalism here resemble its categorical analogue \cite{Chen2021} as much as possible. In Eq.~(\ref{eq:prior_S}), $a$ is a hyperparameter that controls the overall smoothness of the field $\phi$. Smaller values of $a$ correspond to more rugged $\phi$, and larger values of $a$ correspond to more uniform $\phi$. The constant $s$ is arbitrary, but we will see that there is a natural choice for it in a moment. Finally, the parameter $P$ represents the order of the operator $\Delta$ employed to compute the smoothness measure of $\phi$.

Our treatment of ordinal sequence data here is similar to what we do with continuous variables in numerical calculations, in which we first put the problem on a grid and then proceed to compute derivatives of some quantities. However, in our previous work \cite{Chen2021} we found that for functions defined in sequence space it was more meaningful to measure their smoothness by a statistic called ``association'', such as odds, odds-ratio, ratio of odds-ratios, etc. \cite{Bishop1975} than by the ordinary derivatives. 
Thus, we define the operator $\Delta$ in the prior action as follows:
\begin{equation} \label{eq:Delta}
\phi^T \Delta^{(P)} \varphi = \frac{1}{P!} \sum_{i_1, \dots, i_P} \langle D_{i_1} \cdots D_{i_P} \phi, D_{i_1} \cdots D_{i_P} \varphi \rangle,
\end{equation}
where the indices $i_1, \dots, i_P$ must satisfy the conditions $1 \le i_1, \dots, i_P \le \ell$ and $i_1 \ne \cdots \ne i_P$. These conditions immediately imply that $P = 1, 2, \dots, \ell$. Notice that $\phi$ and $\varphi$ are both an $\ell$-dimensional array with $\alpha$ elements in each direction. The symbol $D_i$ denotes taking the discrete difference along the $i$th direction of an array, and the symbol $\langle u, v \rangle$ denotes taking the ``inner product'' of the two arrays $u$ and $v$, that is, multiplying $u$ and $v$ element by element, and then summing up the results. 

So, what this operator $\Delta^{(P)}$ does in the prior is that, given a value of $P$, $1 \le P \le \ell$, it first calculates the discrete difference of the field $\phi$ along all possible combinations of $P$ different directions, and then sums up the squares of all the differences. As an example, consider the toy model in Fig.~\ref{fig:1}, in which $\phi$ is a 3-dimensional array with 3 elements in each direction. Let $P = 2$, for instance, then
\begin{equation} \label{eq:toymodel}
\phi^T \Delta^{(2)} \phi = \| D_1 D_2 \phi \|^2 + \| D_1 D_3 \phi \|^2 + \| D_2 D_3 \phi \|^2,
\end{equation}
where $\| u \|^2 \equiv \langle u, u \rangle$ and we have used the fact that the difference operation is commutative, $D_i D_j = D_j D_i$. Fig.~\ref{fig:2} illustrates the procedure of computing the first term in Eq.~(\ref{eq:toymodel}), $\| D_1 D_2 \phi \|^2$. As shown in the figure, $\phi$ is by construction a $3 \times 3 \times 3$ array. Taking a discrete difference along the second direction, we obtain an array $D_2 \phi$ of dimension $3 \times 2 \times 3$, whose components are differences of adjacent values of $\phi$ in the direction. Taking another discrete difference along the first direction, we obtain another array $D_1 D_2 \phi$ of dimension $2 \times 2 \times 3$, whose components are differences of adjacent values of $D_2 \phi$ in the first direction. The quantity $\| D_1 D_2 \phi \|^2$ is then equal to the sum of the squares of all the components in the array $D_1 D_2 \phi$. The second and third terms in Eq.~(\ref{eq:toymodel}) can be computed in a similar way, and the grand total of the three terms gives the value of $\phi^T \Delta^{(2)} \phi$.

In the example above, each component of the array $D_1 D_2 \phi$ is a hierarchical difference of $\phi$ involving four sequences that form a face in the sequence space. See Fig.~\ref{fig:2}. Since $\phi$ and $Q$ are related via Eq.~(\ref{eq:Q_phi}), hierarchical differences of $\phi$ translate into (negative logarithm of) hierarchical ratios of $Q$. For instance, $(\phi_{221}-\phi_{211})-(\phi_{121}-\phi_{111}) = -\log\{(Q_{221}/Q_{211})/(Q_{121}/Q_{111})\}$. In this case, the hierarchical ratio of $Q$ represents a second-order association (also known as odds ratio) between the first and second sites, conditional on the third site being fixed at 1. In general, the operator $\Delta^{(P)}$ requires taking discrete differences of $\phi$ in $P$ different directions with the remaining directions being the background, and the $2^P$ sequences involved in this procedure form a hypercube of dimension $P$. For example, for $P = 1, 2, 3$, the hypercubes are edges, faces, and cubes, respectively, as shown in Fig.~\ref{fig:1}. Each $P$-dimensional hypercube gives a value of $P$th-order association, which we denote by $\mathcal{A}^{(P)}_k$, where $k$ is the index of the hypercube. This suggests that we can choose the constant $s$ in the prior action to be the number of $P$-dimensional hypercubes embedded in the grid graph and it is equal to $s = {\ell \choose P} (\alpha-1)^P \alpha^{\ell-P}$. Eq.~(\ref{eq:Delta}), with $\phi = \varphi$, can then be rewritten as 
\begin{equation} \label{eq:Delta_A}
\phi^T \Delta^{(P)} \phi = \sum_{k=1}^s \left( \log \mathcal{A}^{(P)}_k \right)^2,
\end{equation}
which is a sum of squared log $P$-associations over all $P$-dimensional hypercubes. The prior distribution of $\phi$, in turn, can be expressed in terms of associations as
\begin{equation} \label{eq:prior_A}
p(\phi | a) \propto \prod_{k=1}^s \exp \left[ -\frac{1}{2} \frac{( \log \mathcal{A}^{(P)}_k )^2}{s/a} \right].
\end{equation}
As we will see below, this expression is very informative and can help us understand some general properties of the solutions.

Having defined the prior distribution of $\phi$, we can obtain its posterior distribution by multiplying the prior by the data likelihood. The resulting posterior distribution of $\phi$ can be expressed as
\begin{equation}
p(\phi | \text{data}, a) \propto e^{-S_a[\phi]},
\end{equation}
where $S_a[\phi]$ is called the posterior action and is given by
\begin{equation} \label{eq:posterior_S}
S_a[\phi] = \frac{a}{2s} \ \! \phi^T \Delta^{(P)} \phi + N R \ \! \phi + N e^{-\phi}.
\end{equation}
Here $N$ is the number of sequences in the dataset, and $R$ is an array having the same shape as $\phi$ and consisting of the observed frequencies. Again, we have used vector notation in Eq.~(\ref{eq:posterior_S}) in order for it to resemble its categorical counterpart \cite{Chen2021}. In the notation introduced in Eq.~(\ref{eq:Delta}), $R \ \! \phi = \langle R, \phi \rangle$ and $e^{-\phi} = \langle \mathbf{1}, e^{-\phi} \rangle$, where $\mathbf{1}$ is an array having the same shape as $e^{-\phi}$ and full of ones. Given a value of $a$, the {\it maximum a posteriori} (MAP) estimate of $\phi$, which corresponds to the mode of the posterior distribution, can be obtained by minimizing the action $S_a[\phi]$. It can be shown that the MAP estimate, denoted $\phi_a$, satisfies the following equation of motion (EOM):
\begin{equation} \label{eq:EOM}
\frac{a}{s} \ \! \Delta^{(P)} \phi_a + N R - N e^{-\phi_a} = 0.
\end{equation}
Varying the value of the hyperparameter $a$, we can obtain a family of MAP estimates, each with a different amount of smoothness. 

From Eq.~(\ref{eq:Delta}), it is clear that the array $\mathbf{1}$ consisting of ones is in the kernel of $\Delta^{(P)}$ for any $P$. Taking the ``inner product'' of $\mathbf{1}$ and Eq.~(\ref{eq:EOM}) leads to $\sum_{i_1 \cdots i_\ell} e^{-(\phi_a)_{i_1 \cdots i_\ell}} = \sum_{i_1 \cdots i_\ell} R_{i_1 \cdots i_\ell} = 1$. Thus, we have $Q_a = e^{-\phi_a}$ from Eq.~(\ref{eq:Q_phi}). In the limit $a = 0$, the EOM implies that $Q_0 = R$, the observed frequency. This can be understood from Eq.~(\ref{eq:prior_A}). We see that when $a = 0$, all the values of $\log \mathcal{A}^{(P)}$ have an infinite variance, meaning that they are totally unconstrained. That is, the prior has no effect at all and the result is completely determined by the data. On the other hand, in the limit $a \rightarrow \infty$, the EOM requires $\Delta^{(P)} \phi_\infty = 0$, meaning that $\phi_\infty$ is restricted to the kernel of $\Delta^{(P)}$. From Eq.~(\ref{eq:prior_A}), we see that in this limit all the values of $\log \mathcal{A}^{(P)}$ are centered at zero with a zero variance, which means that the MAP estimate has vanishing $P$th-order association. In other words, $\phi_\infty$ can only have associations of up to order $P-1$. We can then use ``indicator functions'' of up to $P-1$ sites as the kernel basis of $\Delta^{(P)}$. Denote the $k$-site indicator function by $\mathcal{I}_{s_1, \dots, s_k}^{e_1, \dots, e_k}$, where the subscripts $s_1, \dots, s_k$, each from $1$ to $\ell$, represent the sites, and the superscripts $e_1, \dots, e_k$, each from $1$ to $\alpha$, represent the elements. The indicator function has the same shape as $\phi$, and a component of $\mathcal{I}_{s_1, \dots, s_k}^{e_1, \dots, e_k}$ is equal to one if the elements at sites $s_1, \dots, s_k$ of the corresponding sequence are $e_1, \dots, e_k$, respectively, and is equal to zero otherwise. The set of indicator functions for $k = 1, 2, \dots, P-1$ is able to span the kernel of $\Delta^{(P)}$. Taking the ``inner product'' of $\mathcal{I}_{s_1, \dots, s_k}^{e_1, \dots, e_k}$ and the EOM, the first term vanishes and we have 
\begin{equation} \label{eq:moments}
\mathcal{I}_{s_1, \dots, s_k}^{e_1, \dots, e_k} Q_a = \mathcal{I}_{s_1, \dots, s_k}^{e_1, \dots, e_k} R
\end{equation}
for $k = 1, 2, \dots, P-1$. That is, the MAP estimate $Q_a$ has the same marginal frequencies of up to $P-1$ sites as the data $R$. 

Note that in the limit $a \rightarrow \infty$, the probability distribution is constrained solely by the marginal frequencies. For $P = 2$, the distribution is constrained by the $1$-site marginal frequencies and is thus equivalent to the position weight matrix model. For $P = 3$, the distribution is constrained by the $1$-site and $2$-site marginal frequencies and is thus equivalent to the Potts model. In other words, these traditional models are limiting cases of our formalism. Moreover, in this limit the distribution is given by $Q_\infty = e^{-\phi_\infty}$, where $\phi_\infty = \sum c_{s_1, \dots, s_k}^{e_1, \dots, e_k} \mathcal{I}_{s_1, \dots, s_k}^{e_1, \dots, e_k}$ with some set of coefficients $c_{s_1, \dots, s_k}^{e_1, \dots, e_k}$. These coefficients are chosen so that the distribution satisfies the constraints in Eq.~(\ref{eq:moments}). From a theorem in information theory \cite{Cover2006}, we know that $Q_\infty$ must therefore be the unique maximum entropy distribution. That is, it is the most uniform, and hence has the maximum entropy, distribution satisfying the moments constraints, Eq.~(\ref{eq:moments}). Incidentally, the position weight matrix model and the Potts model are also known as the additive and pairwise maximum entropy (MaxEnt) estimates, respectively. 

In summary, by varying the value of the hyperparameter $a$ from zero to infinity, we can obtain a family of probability distributions from the most rugged data frequency to the most uniform maximum entropy estimate, and each distribution in this family has the same marginal frequencies of up to $P-1$ sites as the data. The next question then is how to find out the optimal value of the hyperparameter. In Bayesian field theory, there are two ways to answer this question: one is by computing Bayesian evidence and the other is by doing cross validation. Here we choose to do $k$-fold cross validation. In $k$-fold cross validation, we first split the data into $k$ even partitions, and then we repeatedly solve for $Q_a$s with $k-1$ partitions and then compute the likelihood of the remaining partition with the resulting $Q_a$s. The optimal value of $a$ is then determined by the likelihood averaged over the $k$ partitions, and we denote the optimal distribution by $Q^*$. This procedure may be time-consuming and computationally demanding, but it is more applicable than computing Bayesian evidence in most practical situations.

%
%
\section{Result \label{Result}}
To demonstrate the utility of our new density estimation method, here we apply it to the aneuploidy data of cancer patients from the TCGA project \cite{Hutter2018}. A cell or an organism is said to be euploid if it has an exact multiple of the haploid number of chromosomes, otherwise it is called aneuploid. For instance, a human cell having three copies of each chromosome is euploid, whereas a human cell having three copies of 22 of the chromosomes but four copies for the remaining chromosome is aneuploid. Aneuploidy is closely related to karyotype. A karyotype records the absolute copy number of each chromosome, while aneuploidy describes by how much the subject deviates from being euploid. An euploid cell or organism may still survive because of balanced gene expression. For an aneuploid cell or organism, unbalanced gene expression most often results in reduced cellular growth rates, but some aneuploid states can in fact accelerate growth \cite{Vasudevan2021}. 

In our previous work \cite{Chen2021}, we analyzed the TCGA aneuploidy data of 10,522 cancer patients across 33 tumor types \cite{Taylor2018}. The dataset contained the aneuploidy of the 22 autosomes, that is, chromosomes other than the sex chromosome. For each patient, we first classified each chromosome as normal, if it was neutral, or abnormal, if it was deleted, amplified, or something more complex. Thus, the aneuploidy data of each patient was simplified to a string of 22 binaries. Then we used SeqDEFT to estimate the probability distribution with the dataset. From a visualization of the resulting probability landscape, we were able to identify some outstanding peaks associated with particular tumor types. One of the peaks corresponded to simultaneous alterations in chromosomes 6, 7, 9, 10, 19, 20, and was found to associate with gliomas. Now we can analyze the case one step further by taking the original four states of aneuploidy into consideration.

We focus on the subset of 1,086 patients that have gliomas \cite{TCGA2008, TCGA2013, TCGA2015}. According to the result of our previous analysis \cite{Chen2021}, we only look at the subset of chromosomes that consists of chromosomes 1, 6, 7, 9, 10, 19, 20. (Chromosome 1 is included because of the 1p/19q codeletion commonly observed in low-grade gliomas \cite{Whitfield2022}.) Each chromosome can take on one of the four states: deleted (D or ``$-$''), neutral (N or ``$\ \ $''), amplified (A or ``$+$''), and complex (C or ``$*$''). The four states are arranged in the order, D, N, A, C, to represent the amount of each chromosome relative to being euploid. This choice of chromosomes and states results in a sequence space of dimension $4^7 =$ 16,384. By comparison, the dataset contains 1,086 sequences ($\mathsf{CNNNNCN}, \mathsf{NNANDNN}, \dots$), among which only 294 sequences are unique. In other words, the observed sequences occupy only $1.79\%$ of the whole sequence space. 

\begin{figure*}[t]
\includegraphics[width=0.329\linewidth]{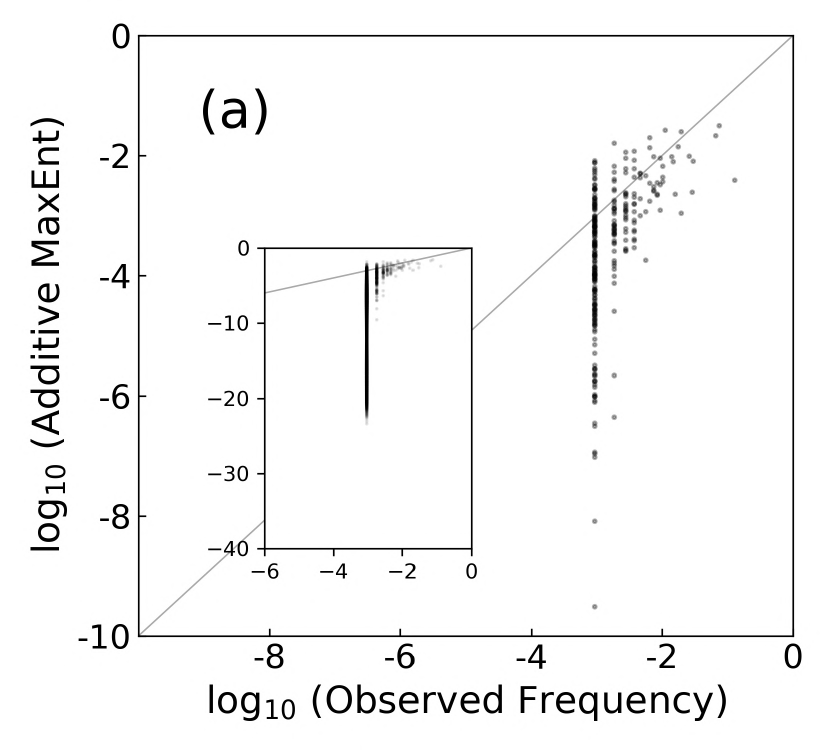}
\includegraphics[width=0.329\linewidth]{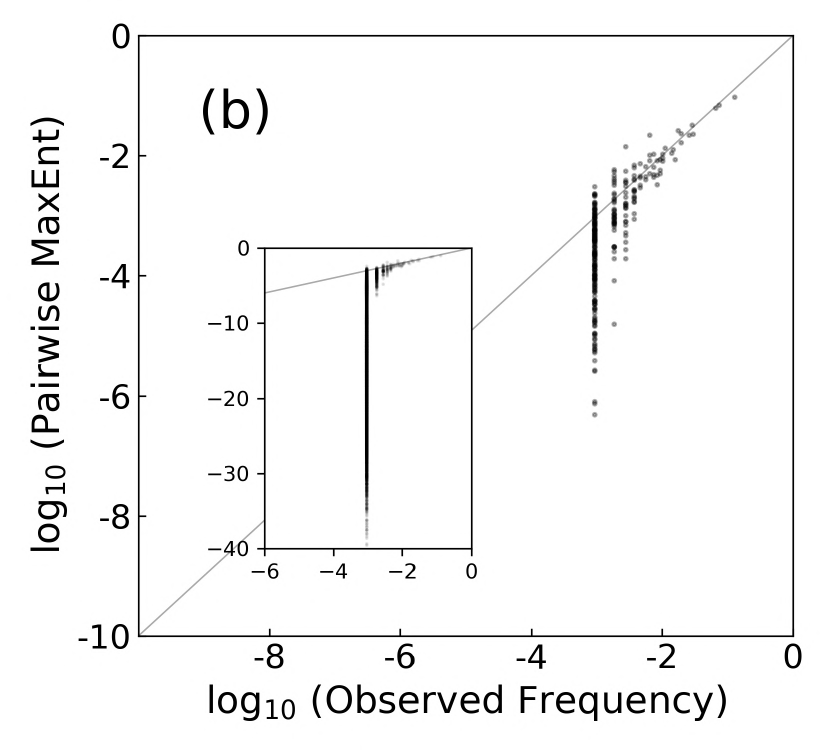}
\includegraphics[width=0.329\linewidth]{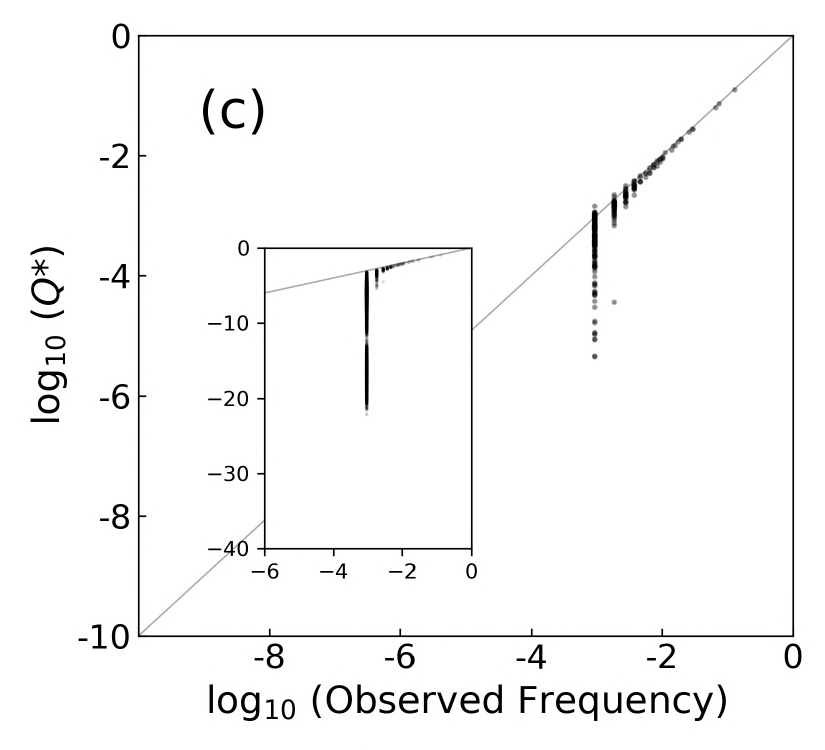}
\caption{Probabilities of the 294 unique sequences in the TCGA gliomas dataset estimated with (a) additive MaxEnt, (b) pairwise MaxEnt, and (c) ordinal SeqDEFT. The insets show the same thing but for all the 16,384 possible sequences in this problem. In order to plot the sequences that were not observed, we add a pseudocount to the dataset. The unobserved sequences are represented by the dots in the last strip in each inset.}
\label{fig:3}
\end{figure*}

%
%
\subsection{Probability distribution}

We use our method with $P = 2$, along with 5-fold cross validation, to estimate the probability distribution. In order to compare, we also fit an additive MaxEnt model and a pairwise MaxEnt model to the same dataset. The choice of the value of the hyperparameter $P$ deserves a few words. In principle, the value of $P$ can be determined entirely from the data. To do so, we compute each family of probability distributions for $P = 1, 2, \dots, \ell$. From the cross-validated log likelihoods of the data, for example, the optimal value of $P$ can be determined. In practice, however, this method is not very practical. This is because sequence data are usually so sparse that the magnitude, or even the existence, of higher-order associations cannot be established. So, we should not choose a high value of $P$. On the other hand, the operator $\Delta^{(P)}$ with $P = 1$ is identical to the graph Laplacian, and thus its function is to make the difference between neighboring sequences as small as possible. But, we have seen in biology that a single point mutation can have a dramatic effect on a sequence. Thus, a value of $P > 1$ should be chosen. Here we find that $P = 2$ works best for the current problem. From the theory we know that the resulting distribution will have the same 1-site marginal frequencies as the data.

The results are shown in Fig.~\ref{fig:3}. From the figure we see that the three models all make reasonably accurate estimates of probabilities for the observed sequences, with the pairwise model providing more precise estimates than the additive model, and our method providing even more precise estimates than the pairwise model. This trend is also reflected in the resulting cross-validated log likelihoods; the additive, pairwise, and ordinal SeqDEFT models give an average cross-validated log likelihood of $-1325$, $-1180$, and $-1140$, respectively. For the unobserved sequences (represented in the last strip in the inset of each subplot), however, it is slightly different. We see that for the unobserved sequences the estimates of the additive model is actually tighter than that of the pairwise model. Intuitively, this is because, although the dataset is large enough to establish the frequency at each site, it is not large enough to establish the correlations between each pair of sites. As a result, the pairwise model gives a wider range of predictions for the unobserved sequences. For ordinal SeqDEFT, on the other hand, in regions where there is no data the distribution is mainly determined by the smoothness operator in the prior. Thus, the distribution therein, which is obtained by pushing the posterior action to zero as much as possible, is similar to the one that is restricted within the kernel of the operator, which for $P = 2$ is an additive model.

We have seen that our new method is capable of inferring the underlying probability distribution given a sample of ordinal sequences. In some cases, the probability distribution itself is what we want. Oftentimes such a probability distribution is then used in follow-up analyses or applications. An example is the generative modeling, in which a probability distribution is used to generate samples having similar features of interest. Below we demonstrate two more things one can do with a probability distribution. One is studying the associations of the sequence sites, the other is probing the global geometry of the probability landscape. We will look at the associations first and then move on to the probability landscape.

%
%
\subsection{Associations of sequence sites}

\begin{figure}[t]
\includegraphics[width=0.8\linewidth]{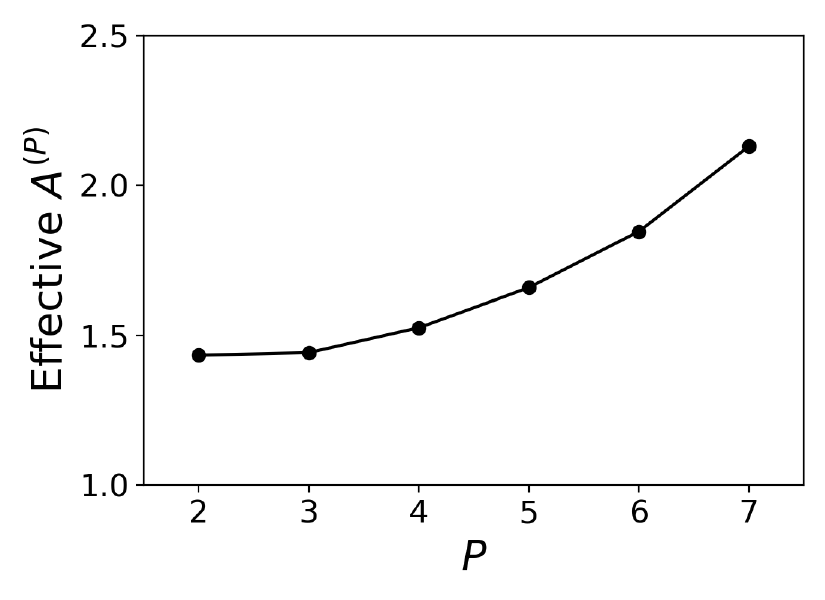}
\caption{Association spectrum inferred from the probability distribution $Q^*$. For each order $P$, the effective magnitude of the association is defined as the exponential of the root-mean-squared value of $\log \mathcal{A}^{(P)}$ computed by using Eq.~(\ref{eq:Delta_A}). The larger the value of $\mathcal{A}^{(P)}$, the stronger the association. An $\mathcal{A}^{(P)}$ of 1.0 means that there is no association at all.}
\label{fig:4}
\end{figure}

From the expression of the smoothness operator in Eq.~(\ref{eq:Delta_A}) we see that sandwiching the operator $\Delta^{(P)}$ with the field $\phi$ representing a distribution results in a sum of squared log $P$-associations over all $P$-dimensional hypercubes. With this quantity we can easily compute the root-mean-squared value of $\log \mathcal{A}^{(P)}$ and use its exponential as the effective magnitude of the $P$th-order association predicted by the distribution in question. Given a distribution, we can repeatedly use Eq.~(\ref{eq:Delta_A}) with $P = 1, 2, \dots, \ell$, to estimate the strength of the association among $P$ sequence sites. The resulting ``association spectrum'' computed with the distribution $Q^*$ we obtained above is shown in Fig.~\ref{fig:4}. Note that the result with $P = 1$ is more like a measure of the overall uniformity of the distribution, rather than an association, so it is not included in the figure. From Fig.~\ref{fig:4} we can see that the sequence sites actually have stronger higher-order associations, although all the associations are only mild. This should not be too surprising since the chromosomes used in the dataset were chosen because of the seeming concurrence of their alterations in gliomas. 

In principle, each value of effective $\mathcal{A}^{(P)}$ in the association spectrum should be computed with a number of log $P$-associations, each corresponding to a particular set of $P$ sites and a certain background. Eq.~(\ref{eq:Delta_A}) provides us a direct formula for the mean squared log $P$-association. We can still compute those log $P$-associations directly and look into them to try to extract more information. When computing a value of, say, log 2-association (i.e., odds-ratio), we need to specify two elements at each of the two sites. For example, $\mathsf{a}$ and $\mathsf{A}$ at one site, and $\mathsf{b}$ and $\mathsf{B}$ at the other. Here $\mathsf{a}$ and $\mathsf{A}$, and $\mathsf{b}$ and $\mathsf{B}$, must each be adjacent in the path graph representing the site. Together, they form four patterns, $\mathsf{ab}, \mathsf{aB}, \mathsf{Ab}, \mathsf{AB}$, at the two sites, which correspond to a face in the grid graph representing the sequence space. See Fig.~\ref{fig:1}. For the sake of  convenience, we can regard the first site as being mutating from $\mathsf{a}$ to $\mathsf{A}$, and the second site from $\mathsf{b}$ to $\mathsf{B}$. This thinking, however, may lead us to a wrong conclusion when interpreting the result. For instance, we may tend to think a strong association between the two sites means that when the first site mutates from $\mathsf{a}$ to $\mathsf{A}$, the second site mutates from $\mathsf{b}$ to $\mathsf{B}$ accordingly, and thus the pattern appearing at the two sites should be $\mathsf{ab}$, $\mathsf{AB}$, or both, rather than $\mathsf{aB}$ or $\mathsf{Ab}$. Although this is indeed one possibility, other possibilities do exist. In fact, one of the other possibilities is a complete opposite of our conclusion, that is, the pattern that appears at the two sites could be $\mathsf{aB}$, $\mathsf{Ab}$, or both, rather than $\mathsf{ab}$ or $\mathsf{AB}$. This problem comes from the fact that a value of log $P$-association is given by a ratio of $2^P$ probabilities, and there are multiple ways to arrange the probabilities so that the log $P$-association has the same magnitude. Thus, there are many scenarios that can result in a strong association between the $P$ sites. The safest way to find out the pattern corresponding to a strong association is therefore by looking into the probabilities of the sequences involved. In our previous example, we may find that $\mathsf{AB}$ has a much larger probability than $\mathsf{ab}, \mathsf{aB}$, and $\mathsf{Ab}$, and thus is the dominant pattern at the two sites. Notice that the pattern found in this manner is merely a ``local rule'', since we are comparing only the four patterns, $\mathsf{ab}, \mathsf{aB}, \mathsf{Ab}$, and $\mathsf{AB}$. If we investigate another face at the same two sites formed by patterns, say, $\mathsf{AB}, \mathsf{AB'}, \mathsf{A'B}$, and $\mathsf{A'B'}$, we may find that a different pattern $\mathsf{A'B'}$, instead of $\mathsf{AB}$, is the most probable one. Local rules like these might be useful in some situations, but a more useful piece of information would be a complete list of the patterns that do appear at the $P$ sites. Such a picture can be patched up with local rules derived from all the $(\alpha-1)^P$ hypercubes. A more straightforward, and perhaps simpler, way to get the picture is by looking into the probabilities of the $\alpha^P$ associated sequences directly. 

Having computed the associations, the next question is how to present the result in an informative way. For the case of $P = 2$, it is an easy task as pairwise associations can be readily represented by using a heatmap. For an example of splice sites in human genome, see reference \cite{Chen2021}. For the cases of $P > 2$, however, presenting those higher-order associations becomes a difficult task. It is clear that the solution should depend on what we want to know about the system. If what we want to know is what combinations of mutations have the strongest associations, and in what backgrounds, we can represent the result in the following manner. First we partition the circumference of a circle into $\ell$ arcs, each arc corresponding to a sequence site, and then we partition each arc into $\alpha - 1$ segments, each segment corresponding to a mutation. Then, any $P$th-order association can be represented by a $P$-sided polygon within the circle connecting the particular mutations involved. This is similar to the ``circos plots'' commonly used in visualizing genomic data \cite{Krzywinski2009}. Meanwhile, if desired, the backgrounds can be represented by a ``sequence logo'' \cite{Tareen2020}. 

\begin{figure*}[t]
\includegraphics[width=0.98\linewidth]{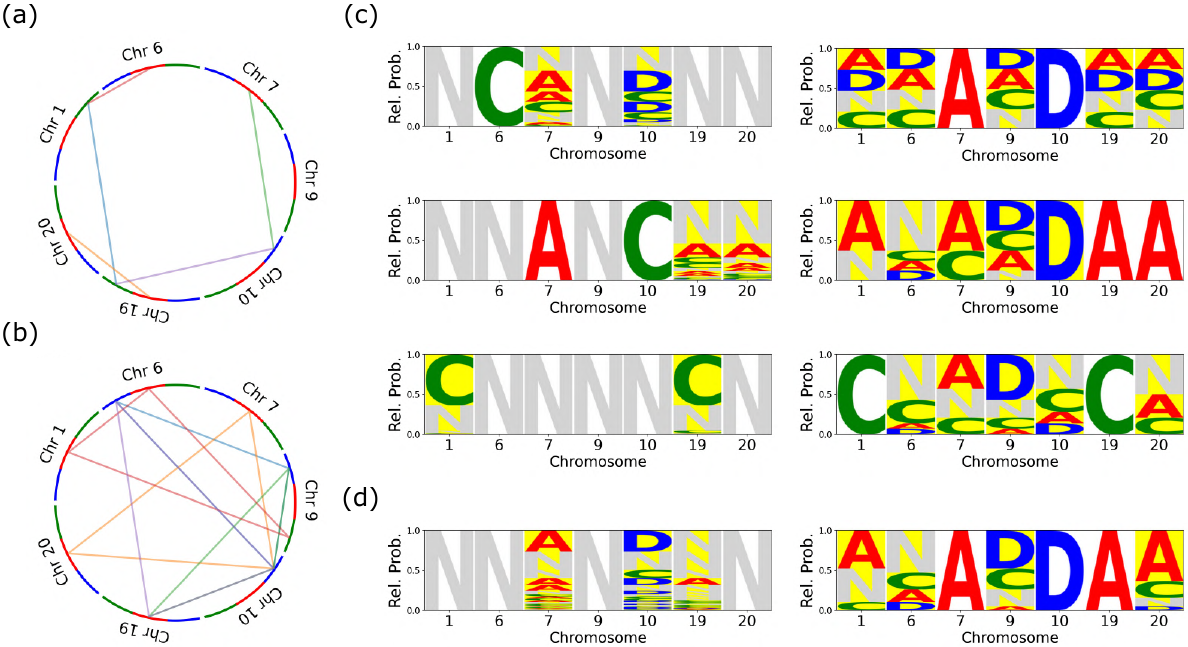}
\caption{(a), (b) Schematic representation of the five strongest associations for $P = 2$ and for $P = 3$ inferred from the probability distribution $Q^*$. In each plot, the circumference of a circle is partitioned into 7 arcs, each corresponding to a chromosome. Each arc is also partitioned into 3 segments, each corresponding to one of the mutations: D $\rightarrow$ N (blue), N $\rightarrow$ A (red), A $\rightarrow$ C (green). Each $P$th-order association is represented by a $P$-sided polygon within the circle. There are five polygons in each plot, colored in blue, orange, green, red, and violet, according to their association strengths in descending order. (c), (d) {\it Left Panel.} Sequence logos that show the patterns appearing at the $P$ sites of interest (highlighted in yellow) conditional on a certain background. The height of each pattern is proportional to its relative probability. {\it Right Panel.} Sequence logos that show the backgrounds (highlighted in yellow) in which a particular pattern tends to appear at the $P$ sites of interest. The height of each letter is proportional to its relative probability.}
\label{fig:5}
\end{figure*}

The five strongest pairwise and three-way associations inferred from the probability distribution $Q^*$ are shown in Figs.~\ref{fig:5}(a) and \ref{fig:5}(b). From Fig.~\ref{fig:5}(a), we can see some well known chromosome abnormalities in gliomas, such as 7${}^{+}$/10${}^{-}$ \cite{Whitfield2022} (conditional on 1/6${}^{*}$/9/19/20) and 19${}^{+}$/20${}^{+}$ \cite{Geisenberger2015} (conditional on 1/6/7${}^{+}$/9/10${}^{*}$). We also see a strong association between 1${}^{*}$ and 19${}^{*}$, conditional on 6/7/9/10/20, which is likely a manifestation of the 1p/19q codeletion \cite{Whitfield2022}. As explained above, to see what is really going on at those pairs of sites, we need to look into the probabilities of the associated sequences. Here we examine all the patterns that are allowed at the two sites, not just the patterns formed by the mutations, to get a more complete picture. The results are displayed as sequence logos in the left panel of Fig.~\ref{fig:5}(c). We can see that the patterns 7${}^{+}$/10${}$, 19${}^{+}$/20${}^{+}$, and 1${}^{*}$/19${}^{*}$ are indeed the most prominent pattern at the two sites when both sites mutate. Note that, for chromosomes 1 and 19, the pattern 1${}^{*}$/19${}^{*}$ is more probable than the wild type. 

These pairwise associations, 7${}^{+}$/10${}^{-}$, 19${}^{+}$/20${}^{+}$, and 1${}^{*}$/19${}^{*}$, have an effective strength (averaged over all backgrounds) of 2.52, 2.02, and 1.58, respectively, which are greater than 99$\%$, 98$\%$, and 88$\%$ of all possible pairwise associations. We plot the backgrounds in which 7${}^{+}$/10${}^{-}$, 19${}^{+}$/20${}^{+}$, or 1${}^{*}$/19${}^{*}$ is the most probable pattern at the two sites as sequence logos in the right panel of Fig.~\ref{fig:5}(c). The backgrounds for each pattern are found in the following manner. For each background, we first find the probabilities of all the patterns ($\mathsf{DD, DN, ..., CC}$) at the two sites. This gives us a vector of 16 components, and we normalize it to 1. Then, a ``similarity score'' can be obtained by taking the inner product of this vector with another vector constructed in the same manner with the pattern of interest. A background is selected if the similarity score is greater than 0.5. We find 950, 8, and 14 backgrounds for 7${}^{+}$/10${}^{-}$, 19${}^{+}$/20${}^{+}$, and 1${}^{*}$/19${}^{*}$, respectively. From the figure, we can see that the backgrounds for 7${}^{+}$/10${}^{-}$ have no specificity, meaning that each state appears with an equal probability at the background sites. By contrast, the backgrounds for 19${}^{+}$/20${}^{+}$ and 1${}^{*}$/19${}^{*}$ show a mild to medium specificity. 

Higher-order associations have been rarely identified in clinical studies of gliomas. An alluded third-order association involves 7${}^{+}$/10${}^{-}$/19${}^{+}$ \cite{Cohen2015} which, however, does not show up in Fig.~\ref{fig:5}(b). We compute the probabilities of all the patterns at the three sites with a neutral background. The result is shown in the left panel of Fig.~\ref{fig:5}(d), from which we can see that 7${}^{+}$/10${}^{-}$/19${}^{+}$ is indeed the most probable pattern when the three sites all mutate. On the other hand, the effective strength (averaged over all backgrounds) of this abnormality is inferred to be 1.56, which is greater than 81$\%$ of all possible third-order associations. Moreover, we find the backgrounds in which 7${}^{+}$/10${}^{-}$/19${}^{+}$ takes place in the same manner as above. The 19 backgrounds found are plotted in the right panel of Fig.~\ref{fig:5}(d), which also show a mild specificity. 

We have demonstrated that the association analysis presented above can help us find out the patterns that tend to appear at some sites in a certain background. It can also help us find out in what backgrounds a particular pattern is more likely to appear at some sites. This information, in turn, can tell us which mutations tend to take place together. These rules, or ``biological grammar'', dictate how different sequence sites should cooperate together in order to make a sequence ``visible'', and thus may be useful in the exploration of underlying physical or biological mechanisms. Here we analyze only the pairwise and three-way associations. Higher-order analysis can be carried out in the same manner.

\begin{figure}[!ht]
\includegraphics[width=1.0\linewidth]{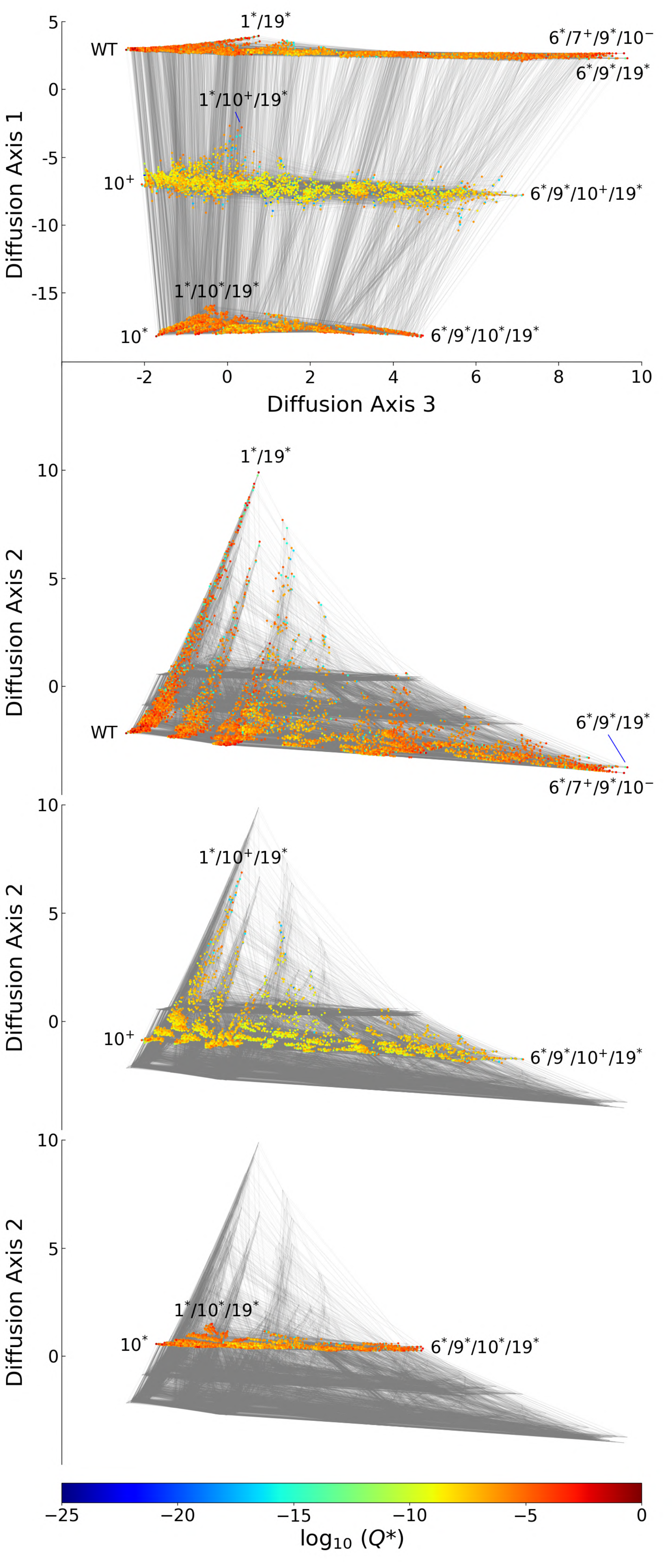}
\caption{Visualization of the probability distribution $Q^*$. Each node represents a sequence and is colored by its inferred probability, and each edge represents a single point mutation. The sequences form three groups, conditional on 10/10${}^{-}$, 10${}^{+}$, or 10${}^{*}$, respectively, which are plotted separately in the lower panel for the sake of visibility. WT, wild type.}
\label{fig:6}
\end{figure}

%
%
\subsection{Global geometry of probability landscape}

Now we turn to investigate the probability landscape of the distribution $Q^*$. The distribution $Q^*$ is a very high dimensional object. Thus, in order to visualize it, we resort to a dimensionality reduction technique based on an evolutionary model \cite{McCandlish2011}. In this method, the distribution $Q^*$ is regarded as being the result of an evolutionary process. This evolutionary process can be represented by a reversible Markov chain, whose rate matrix is given by
\begin{equation}
T_{ij} = \frac{\log Q^*_j - \log Q^*_i}{1 - e^{-(\log Q^*_j - \log Q^*_i)}},
\end{equation}
where $i$ and $j$ denote mutationally adjacent sequences, and the diagonal element $T_{ii}$ is chosen so that the row sums of the matrix $T$ are all zero. Order the eigenvalues of the matrix $-T$ in such a way that $0 = \lambda_1 < \lambda_2 \le \lambda_3 \le \cdots \le \lambda_{\alpha^\ell}$. A $d$-dimensional visualization of the Markov chain, and hence the distribution $Q^*$, can be obtained by plotting each sequence $i$ with coordinates $\sqrt{1/\lambda_2} \ r^{(2)}_i, \cdots, \sqrt{1/\lambda_{d+1}} \ r^{(d+1)}_i$, where $r^{(k)}$ is the right eigenvector of $-T$ corresponding to the eigenvalue $\lambda_k$. It can be shown that the reciprocal of the square root of the $k$th eigenvalue, $\sqrt{1/\lambda_k}$, is proportional to the variance explained in the direction of the $k$th eigenvector. Moreover, the squared distance between two sequences $i$ and $j$ in the visualization approximates the expected ``commute time'' between them (i.e., the time it takes to go from $i$ to $j$ and then back to $i$). For more details, see reference \cite{McCandlish2011} and the supplement to \cite{Chen2021}.

\begin{figure*}[t]
\includegraphics[width=0.47\linewidth]{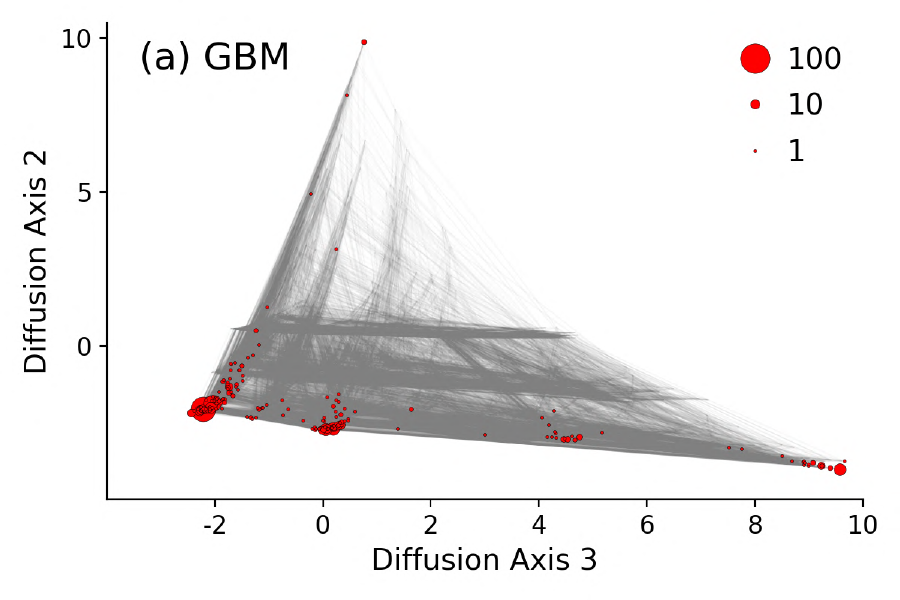}
\hspace{0.8cm}
\includegraphics[width=0.47\linewidth]{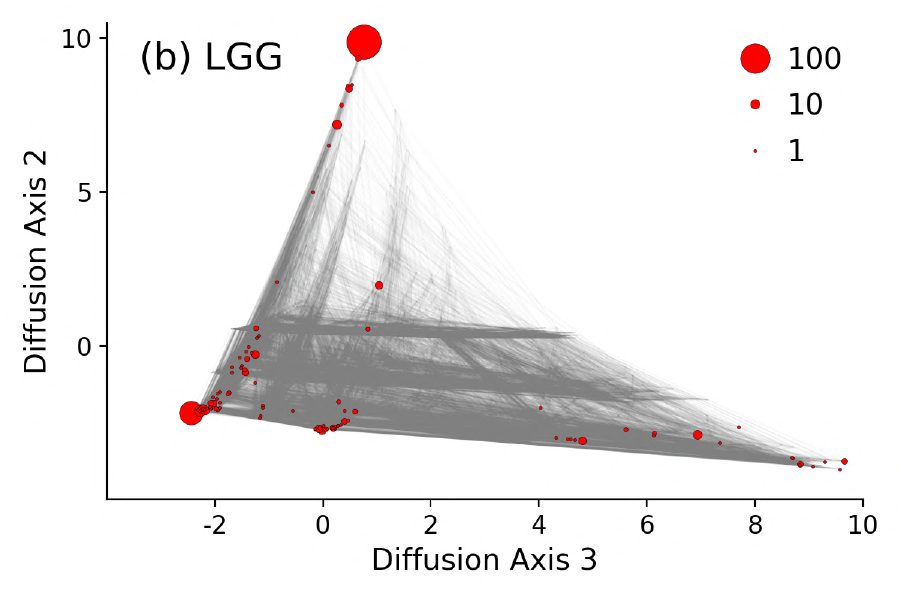}
\caption{Number of patients with a particular pattern of the seven chromosomes from the TCGA dataset. Note that the dataset is based on the old definitions of cancer types. Only the patients belonging to the group with higher values on Diffusion Axis 1 in Fig.~\ref{fig:6} are counted. The patients are further separated into two subgroups, one consisting of those patients annotated as having glioblastoma (GBM, left panel) and the other with patients annotating as having low-grade gliomas (LGG, right panel). The size of each dot is proportional to the number of patients having the corresponding pattern.}
\label{fig:7}
\end{figure*}

Fig.~\ref{fig:6} shows a visualization of the probability distribution $Q^*$ using the method described above with the first three rescaled eigenvectors. These rescaled eigenvectors are called ``diffusion axes'' since they capture the slow modes of the underlying evolutionary process \cite{McCandlish2011}. Each node in the visualization represents a sequence and is colored by its inferred probability, and each edge represents a single point mutation. We can see that the sequences are divided into three distinct groups by Diffusion Axis 1. A closer inspection reveals that higher, medium, and lower values on Diffusion Axis 1 correspond to sequences with chromosome 10 being deleted or neutral, amplified, and complex, respectively. This segregation may seem strange at first since each chromosome is allowed to be in any one of the four states. To understand why this happens, we note that the chromosome abnormality 10${}^{-}$ is very common in gliomas. In fact, in the TCGA dataset 10${}^{-}$ is more frequent than the wild type. On the other hand, the abnormality 10${}^{+}$ is quite rare in the dataset. Thus, from an evolutionary point of view, a population starting as a wild type is much more likely to evolve to 10${}^{-}$ than to 10${}^{+}$. In other words, the commute time between wild type and 10${}^{-}$ is much shorter than the commute time between wild type and 10${}^{+}$. In the visualization the sequences with chromosome 10 being deleted and the sequences with chromosome 10 being neutral form two individual groups; however, because of their very short commute time, they appear to be overlapping with each other and are far from the group of sequences with 10${}^{+}$. The chromosome abnormality 10${}^{*}$ is also much more frequent than 10${}^{+}$, which separates the two groups of sequences by a large distance in the visualization as well. As a result, the sequences are segregated into three distinct groups. 

Diffusion Axis 3 is found to play a similar role as Diffusion Axis 1 in the visualization, with the controlling chromosome now being chromosome 6. More specifically, the region with lower values on Diffusion Axis 3 corresponds to a mix of sequences with chromosome 6 being deleted or neutral, whereas the regions with medium and higher values on Diffusion Axis 3 correspond to sequences with chromosome 6 being amplified and complex, respectively. Although the segregation along Diffusion Axis 3 is not as prominent as that along Diffusion Axis 1, it can still be discerned by the various shades of orange along the axis. 

Unlike Diffusion Axes 1 and 3, Diffusion Axis 2 does not have a clear connection with the states of a certain chromosome. Nevertheless, the sequences form several stripes that stretch to different extents along Diffusion Axis 2. We find that the farthest nodes of each stripe always possess the pattern 1${}^{*}$/19${}^{*}$ with some background, whereas the middle portion of each stripe consists exclusively of sequences either having 1${}^{*}$/19${}^{+}$ or having 1${}^{+}$/19${}^{*}$, and this trend vanishes as we move further toward the head of the stripe. What distinguishes the stripes? It is found that the sequences forming the leftmost stripe in the first group all have chromosome 9 being neutral or deleted, and the sequences forming the next stripe therein all have chromosome 9 being amplified. However, distinctions between other stripes are hard to find as the boundaries between them are blurry. In addition, we also find that in each group sequences having 7${}^{+}$ tend to gather around the bottom (i.e., regions with lower values on Diffusion Axis 2), although some of them may scatter deep into a stripe.

From the global geometry of the probability landscape, Fig.~\ref{fig:6}, we can imagine an evolutionary process as follows. A population starts as a wild type, with all the chromosomes being neutral. As the population wanders around the sequence space, it has a much higher probability of staying in the original group than jumping to the group with 10${}^{+}$. Even if the population does jump to the group with 10${}^{+}$, it will soon jump back to the original group or jump to the group with 10${}^{*}$. Let us focus on what is likely to happen to the population, assuming that it stays in the original group. Because the group with 10${}^{-}$ overlaps thoroughly with the group with 10 neutral, the population is very likely to pick up the abnormality 10${}^{-}$. In addition, since the wild type is surrounded by sequences having 7${}^{+}$ (which tend to gather near the bottom of the group), the population is also very likely to pick up the abnormality. This combined pattern 7${}^{+}$/10${}^{-}$ is a signature of glioblastoma \cite{Whitfield2022}. Sequences having both 7${}^{+}$ and 10${}^{-}$ abnormalities form the routes from wild type to the bottom-right corner along Diffusion Axis 3. Our population may also take on alternative routes in the direction of Diffusion Axis 2. In doing so, it picks up either 1${}^{*}$ or 19${}^{*}$, and eventually possesses the pattern 1${}^{*}$/19${}^{*}$, which in our analysis is characteristic of low-grade gliomas \cite{Whitfield2022}. Another possibility is that the population may pick up the abnormality 7${}^{+}$/10${}^{-}$ on its way toward the end of a stripe. This corresponds to the clinical scenario that some low-grade gliomas may progress to glioblastoma \cite{TCGA2015}. 

Fig.~\ref{fig:7} shows how many times a particular chromosome abnormality with 10 or 10${}^{-}$ is observed in the TCGA dataset. From the figure we can see that the abnormalities with glioblastoma being the diagnosis scatter primarily around the bottom, whereas the abnormalities with low-grade gliomas being the diagnosis tend to pile up at the upper corner. This picture is consistent with what we have inferred from an imaginary evolutionary process taking place in the probability landscape shown in Fig.~\ref{fig:6}. 

%
%
\section{Summary and Discussion \label{Summary and Discussion}}
Within the framework of Bayesian field theory, we developed a novel approach to estimate the probability distribution from which a sample of ordinal biological sequences were drawn. This approach was based on the idea of representing the sequence space by a suitable graph. We found that for ordinal sequences a grid graph, which is a Cartesian product of path graphs, was a suitable choice as it can take into account the ordering of sequence elements. In such a representation, we then constructed an operator which can compute the associations among sequence sites of any order, and these associations were used to form a smoothness measure. This operator played a role like a ``regularizer'' in machine learning, allowing us to control the smoothness of the resulting distribution by tuning the value of a hyperparameter, and the optimal value of the hyperparameter was determined via $k$-fold cross validation. We applied this approach to the TCGA aneuploidy data of gliomas, and we found that the optimal distribution inferred with this approach had a better performance than the parametric additive and pairwise MaxEnt models. 

We also did two follow-up analyses with the optimal distribution. Firstly, we computed an association spectrum that revealed the association strengths of the sequence sites of all orders. We also looked into the associations of second and third orders to search for events of interest, for example, what combinations of mutations tended to occur simultaneously and in what backgrounds. This analysis provided us a way to figure out the underlying biological grammar for the system. Indeed, we were able to recover several well known chromosome abnormalities in gliomas. Secondly, we used a dimensionality reduction technique based on an evolutionary model to visualize the probability landscape. We saw that the global geometry of the probability landscape reflected some prominent features of the system. Moreover, we investigated how a population starting as the wild type might evolve to other states and eventually ended up in glioblastoma or low-grade gliomas. Looking at the system from this perspective helped us understand the underlying biological mechanisms.

Finally, we conclude with an issue which we believe is worth further investigation. In our previous work of density estimation on categorical sequence data \cite{Chen2021}, the sequence space was represented by a Hamming graph, namely, a Cartesian product of complete graphs. We found that there existed a simple and elegant expression of the smoothness operator $\Delta^{(P)}$ in terms of the Laplacian matrix of the Hamming graph and its eigenvalues. Specifically, for categorical sequences of length $\ell$ and $\alpha$ elements, $\Delta^{(P)}$ can be expressed as
\begin{equation} \label{eq:Delta_decomp}
\Delta^{(P)} = \frac{1}{P!} \prod_{k=0}^{P-1} (L - k \alpha I),
\end{equation}
where $L$ is the Laplacian matrix of the graph, $I$ is the identity matrix, and $k \alpha$, for $k = 0, 1, 2, \dots, \ell$, are the eigenvalues of $L$. This operator can compute associations for all possible combinations of $P$ different sites, as in Eq.~(\ref{eq:Delta_A}). This expression not only suggests a more efficient way to implement the computation of associations but also gives us an insight into the sequence space decomposition. Unfortunately, when developing the theory of this work, we found that this expression no longer worked out and we had to stick with a raw definition of $\Delta^{(P)}$ as in Eq.~(\ref{eq:Delta}). The failure of Eq.~(\ref{eq:Delta_decomp}) for grid graphs can be easily seen by trying with a toy model. But why does it not work out? Does the expression exist only for Hamming graphs? We think it is worthy to investigate if a decomposition similar to Eq.~(\ref{eq:Delta_decomp}), possibly with different entities or even in a different mathematical form, exists for other composite graphs. If not, what properties disable a graph from having such a decomposition? We believe answering these questions will be of interest to graph theory in general and sequence data analysis in particular.

Our implementation of ordinal SeqDEFT is available at \url{https://github.com/wcchen-ccu/OrdinalSeqDEFT}. The TCGA dataset of aneuploidy in human cancer is from reference \cite{Taylor2018} and is available as part of the supplementary material at \url{https://doi.org/10.1016/j.ccell.2018.03.007}. 

\begin{acknowledgments}
W.C.C. acknowledges support from the National Science and Technology Council of Taiwan, R.O.C., under Grant No. NSTC 111-2112-M-194-008-MY3 and additional funding from National Chung Cheng University. D.M.M. acknowledges support from NIH grant R35GM133613, an Alfred P. Sloan research fellowship, and additional support from the Simons Center for Quantitative Biology at Cold Spring Harbor Laboratory.
\end{acknowledgments}

\bibliography{main}

\end{document}